ISTITUTO NAZIONALE DI FISICA NUCLEARE

Laboratori Nazionali di Frascati

______________________________________________

# POSEYDON – CONVERTING THE DAΦNE COLLIDER INTO A DOUBLE POSITRON FACILITY: A HIGH DUTY-CYCLE PULSE STRETCHER AND A STORAGE RING

Paolo Valente[1]

[1] *INFN, Sezione di Roma, P.le Aldo Moro, 2, I-00185 Roma, Italy.*

## Abstract

This project proposes to reuse the DAΦNE accelerator complex for producing a high intensity (up to $10^{10}$), high-quality beam of high-energy (up to 500 MeV) positrons for HEP experiments, mainly – but not only – motivated by light dark particles searches. Such a facility would provide a unique source of ultra-relativistic, narrow-band and low-emittance positrons, with a high duty factor, without employing a cold technology, that would be an ideal facility for exploring the existence of light dark matter particles, produced in positron-on-target annihilations into a photon+missing mass, and using the bump-hunt technique. The PADME experiment, that will use the extracted beam from the DAΦNE BTF, is indeed limited by the low duty-factor ($10^{-5}$=200 ns/20 ms).

The idea is to use a variant of the third of integer resonant extraction, with the aim of getting a $<10^{-6}$ m·rad emittance and, at the same time, tailoring the scheme to the peculiar optics of the DAΦNE machine. In alternative, the possibility of kicking the positrons by means of channelling effects in crystals can be evaluated. This would not only increase the extraction efficiency but also improve the beam quality, thanks to the high collimation of channelled particles. This is challenging for < GeV leptons, and in particular this would be the first positron beam obtained with crystal-assisted extraction (generally limited to protons and ions).

The availability of an intense extracted positron beam with a tuneable pulse length will also enable other applications, ranging from radiation production from crystal undulators to irradiation for aerospace industry.

The second ring can be used for storing positrons accelerated by the LINAC, both for producing synchrotron radiation (reversing the polarity of the ring currently used for electrons) and for machine studies with positively charged particles, like for instance instabilities driven by the electron cloud effect.

# Table of contents

# 1. Motivations

## 1.1 Introduction

The DAΦNE double-ring collider [1] in Frascati laboratory (LNF), running at the Φ meson resonance center of mass energy of ~1 GeV, will soon complete the Kaon physics program [2]. The general layout of the complex, including facilities like synchrotron radiation laboratories and beam-test is shown in Fig. 1. One possibility for extending the life of such a precious machine would be to use the stored electrons (or positrons, as in the case of the CESR storage ring at the Cornell University [3]) for producing synchrotron light. However, the availability of a rare resource like a powerful source of high-energy positrons also opens new research opportunities.

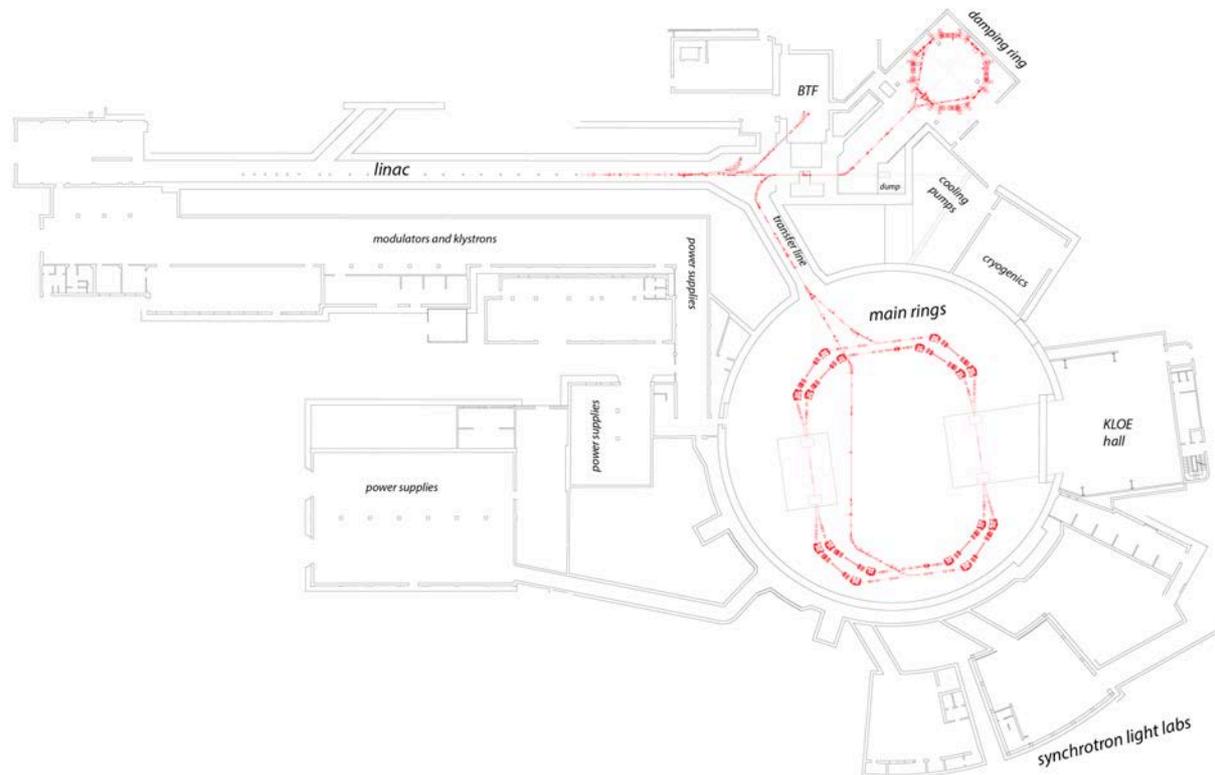

**Figure 1.** *Layout of the DAΦNE complex at the Frascati laboratory.*

Ultra-relativistic positron (>>MeV) are generally produced by high-current electron radio-frequency (RF) LINAC's, with repetition rate up to hundred Hz, and beam pulse duration in a very wide range, from below ps (in the case of photo-injectors) up to few μs range (in case of thermo-ionic guns and uncompressed RF power); charge usually ranges from pC to few nC per pulse. Positrons are generally produced by Bremsstrahlung onto high-$Z$ targets with a very high emittance, and then focused by strong magnetic fields. An alternative production method is by means of the intense photon flux generated by electrons in an undulator. The relatively short number of accelerators running with positrons is further reduced when considering externally extracted beams, either directly from the LINAC or from the storage ring, as briefly reviewed in the following sections. Un-separated positrons beams can be produced by proton primaries, like in the CERN H4 line, delivering up to 200 GeV positrons from the protons extracted from the SPS on one of the North Area targets.

Finally, a ring operating with a high current of positrons, is an ideal laboratory for studying electron cloud effects, which is an important limiting factor for present and future high intensity proton and e+ e− machines.

## 1.2 Positrons at DAΦNE and BTF

In the DAΦNE LINAC [4] positrons are produced by striking ~5.5A, ~200 MeV electrons on a ~2-radiation lengths Tungsten-Rhenium target (positron converter, see Fig. 2). Electrons are produced by the thermo-ionic gun, bunched (prebuncher and buncher cavities, PB and B in Fig. 2), and accelerated in the first five accelerating sections (E1-E5). Downstream of the converter, strong solenoids collect emitted positrons, which are captured by a high-gradient section (capture section, CS), followed by a standard one (P1), and are then separated from electrons by a four-dipoles achromatic bump. Half of the RF power, provided by two modulators and 45 MWp klystrons is used in this first half of the LINAC. The remaining 8 accelerating sections, fed by two additional RF stations (identical the former ones), bring positrons to a maximum energy of 530-550 MeV. Typically, with 10 ns long pulses, up to 1 nC charge is produced.

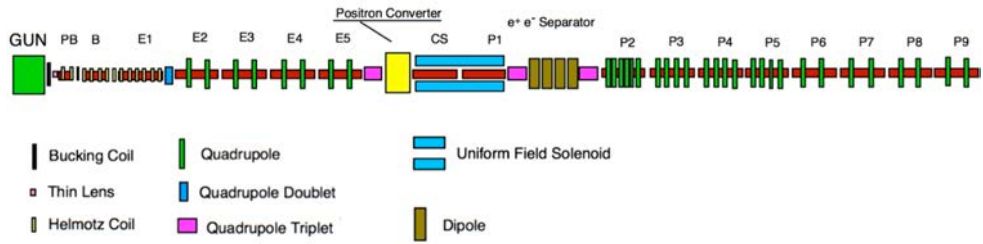

**Figure 2.** *Logical scheme of the DAΦNE LINAC and focussing system.*

Alternatively, positrons can be also produced in the Beam-Test Facility (BTF) transfer line [5], by selecting positive secondary particles emerging from the beam-attenuating target. Electrons or positrons pulses from the LINAC can be diverted to the BTF line by means of a pulsed 3° dipole; during standard running of the collider, all LINAC pulses not used for injection in the DAΦNE rings are available for the BTF (excluding pulses used for monitoring purposes, once or twice per second).

In this case the energy of the positrons can be adjusted (by means of a dipole and a collimator system) from the primary LINAC energy down to few tens of MeV, but a significantly lower intensity beam can be produced (from "single particle" regime to $10^6$/s, depending on the energy). The layout of the transfer-lines for delivering electrons and positrons from the LINAC to the BTF and to DAFNE rings is shown in Fig. 3.

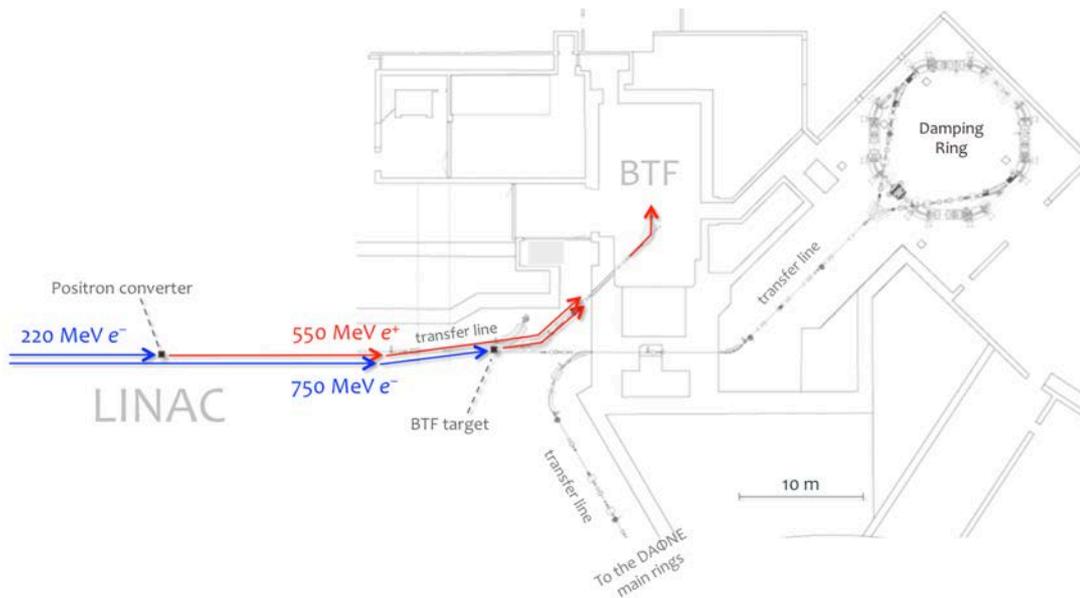

**Figure 3.** *The BTF transfer line and the two possible positron beams: LINAC primary positrons from the converter, or secondary beam from the BTF target.*

## 1.3 Positrons for dark sector experiments

High-energy, high-intensity positrons are of prime importance for fundamental physics, in particular, *e+* fixed target annihilations for dark sector particle searches [6,7] can profit of a quasi-continuous, high-energy positron beam. The first experiment searching for light dark matter particles with the missing mass technique in positron fixed-target annihilations will be PADME [8-10] using the BTF improved line [11], using $10^4$–$10^5$ positron pulses, of 150-200 ns length and 550 MeV energy, on a very thin, active diamond target. A first run of 4-6 months should allow reaching $10^{13}$ positrons on target, corresponding already to a sensitivity of $\varepsilon \sim 10^{-3}$, i.e. already in the interesting band favoured by the muon *g*-2 anomaly, as shown in Fig. 4; the potential of PADME for a integrated statistics of $4\cdot10^{13}$ positrons is also shown.

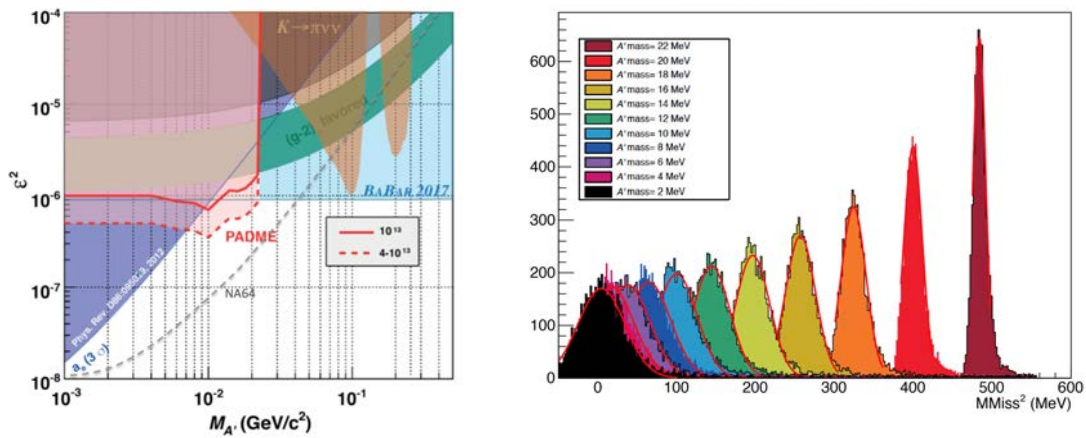

**Figure 4.** *Left: expected sensitivity of the PADME experiment for $10^{13}$ (full line) and $4\cdot10^{13}$ (dashed line) positrons on target; the exclusions in invisible decays are also shown (see text). Right: missing mass distributions for dark photons of masses spanning from 2 to 22 MeV/$c^2$, simulated for the PADME experiment.*

The PADME experiment will give the unprecedented possibility of positively looking for a peak in the missing mass spectrum (Fig. 4, right) in events $e+e- \rightarrow \gamma$ + invisible particles, in a clean and model-independent way, up to ~24 MeV/$c^2$ mass, thanks to a finely segmented, high-resolution inorganic crystals calorimeter, and a number of veto detectors [12] (see Fig. 5). This model-independent technique is sensitive not only to dark vector (dark photon), but also to more exotic dark sector candidates, like the proto-phobic boson [13] claimed to be responsible of the anomaly in the internal pair angular distribution in $^8$Be radiative transitions [14], or the axion-like particles [15].

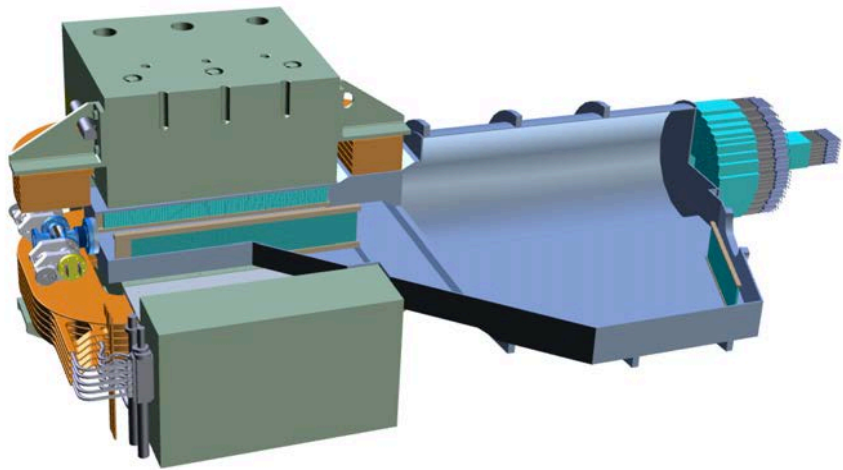

**Figure 5**. *PADME setup view, showing the target area, the dipole magnet with embedded scintillators as positron veto, the vacuum vessel with the additional scintillator detector and the main cylindrical BGO crystal calorimeter, followed by the fast small angle photon detector.*

The main limitations related to the positron source for this kind of experiments [16] are the time structure of beam pulse, due to the pile-up of close events in the detectors, limiting the maximum intensity (and hence the luminosity), and the divergence and energy spread of the beam, since the precise knowledge of the positron momentum vector impacts the missing mass resolution. Another limiting factor in the fixed-target approach is of course the maximum beam energy $E_+$, giving access to a low mass range in the center of mass ($M^2 < 2m_e E_+$).

The duty-factor is generally low in the case of warm LINAC's, on the contrary, continuous-wave beams are produced using superconducting cavities, but with much higher cost and complexity. The production of intense positron beams (both polarized and unpolarized) is very interesting for fundamental physics (like the above-mentioned dark sector experiments), so that currently is being studied at large accelerator facilities using cold technology, like MESA in Mainz [17], LCLS-II at SLAC and CEBAF at JLAB [18], and is under consideration also for the proposed electron-ion collider.

In the case of DAΦNE the duty factor is limited to ~$10^{-5}$, since the 50 Hz LINAC is capable of accelerating pulses up to few hundreds of ns [19] due to the compression of the RF power by means of the SLAC Energy Doubler device (SLED), which enables to reach a maximum energy of 750 MeV/550 MeV for electrons/positrons with the relatively low power. This limits the possibility of increasing the beam macro-pulse length, in order to get a higher luminosity, while keeping the pile-up probability in the recoiling photon detector to a manageable level.

A possibility at the BTF would be to use the LINAC without the SLED compression: thanks to the new electron gun pulsing system [20] beam pulses can be produced and accelerated up to the maximum klystron pulse length of 4.5 μs, however in this case the maximum energy is lowered to approximately one half, i.e. up to 250–280 MeV for positrons.

An alternative approach, proposed for instance at the VEPP-3 $e+e-$ machine in Novosibirsk, is to use the circulating positron beam, colliding it with an internal thin target [21]. The corresponding luminosity would be in this case affected by the need of introducing a small perturbation on the stored positron bunches, in particular using a low density target (gaseous Hydrogen typically, since low-Z is also required for maximizing the annihilation cross-section with respect to the Bremsstrahlung one), or better a jet-target. Another issue of the internal target approach is the limitation on the recoiling photon acceptance, due to the unavoidable machine elements at small angle (with respect to the positron direction), relevant for higher energy photons corresponding to the lower end of the dark photon mass range.

## 1.4 Positrons for channelling experiments

Another interesting application of a high-energy, high-quality positron beam is in the generation of coherent radiation, by means of crystal undulators [22], exploiting channelling effects of the positron in crystalline structures [23], since it is the lightest positively charged particle. The great advantage over other MeV-range, narrow-band photon sources, like inverse Compton scattering on a high-energy electron beam, would be the simplicity and efficiency of the radiation generation, since it does not require a short-pulse and powerful laser source, and the beam charge can be increased by using a simpler thermo-ionic gun.

At present, crystal undulators are routinely used for producing photons by means of coherent Bremsstrahlung of high-intensity, sub-GeV electron beams, like in MAMI (Germany), MAXLAB (Sweden) and JLAB (USA), in the region of ~100 keV. Channelling radiation would allow producing ~MeV, high-quality photon beams starting from the same energy, as shown from Fig. 6, adapted from Ref. [24].

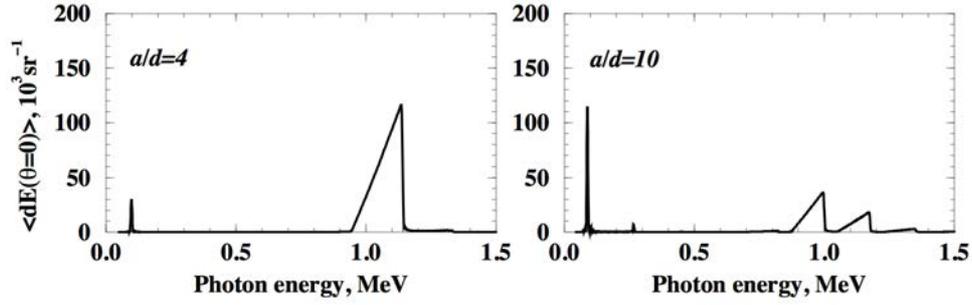

**Figure 6.** *Spectral distribution of radiation emitted in the forward direction for 500 MeV positrons channelling in 0.35 mm thick Silicon along the (110) crystallographic planes, for two different values of the ratio a/d, where a is the amplitude of the bending and d is the planar distance (from Ref. [24]).*

## 2. Project description

*2.1 Main objective*

The first objective of this proposal is the design, optimization and realization of an intense, high-duty cycle, narrow-band and low-emittance ultra-relativistic positron source, for a wide range of scientific and technological applications, realized using the available accelerator complex of DAΦNE, with some improvements and modifications.

More specifically the aim is to efficiently stretch the short pulses of the positron beam produced by the high-current LINAC, by means of one of the two rings, extracting it to a dedicated transfer line, also allowing to independently use the second ring (both for the production of synchrotron radiation, already serving the existing lines of DAΦNE-Light laboratory, and for dedicated single-beam machine studies), thus fully optimizing the use of the existing infrastructure.

An extraction scheme capable of producing a high-quality positron beam without using the wiggler magnets installed in the rings, otherwise necessary for high-luminosity *e+e-* collisions, would greatly reduce the electricity needed for the operation: since wigglers are the most energy-hungry elements in DAΦNE, the overall power of 3.5 MW would be reduced by more then a factor two [25], even running with two beams (and considering that one out of eight wigglers is used for one of the synchrotron light lines).

In order to define the project details, the main parameters for the stretched beam should be defined:

- Energy: the maximum achievable energy is fixed not only by the original design of the rings, relatively short (~100 m) so that the ~1.4 m long dipoles can be pushed up to 750–800 MeV before reaching iron saturation, but more importantly from the maximum positron energy from the LINAC, presently limited to 550 ~MeV.
- Energy spread: improving the 1% level achieved using the BTF line, would impact mainly the potential of fixed-target experiments, where the precise definition of the kinematics is very important, but would be also beneficial to all other potential applications of the high-energy positron source.
- Emittance: it defines a good quality of the beam in terms of angular and spatial spread; a target level is $10^{-6}$ m·rad at 500 MeV, and must be preserved by the extraction beam-line. This is an order of magnitude improvement with respect to the DAΦNE LINAC positron beam.
- Duty-cycle: possibly the most important parameter in order to have a high average intensity source, defined as the fraction of time of the beam spill/total time. Of course high peak intensity can be obtained using intense and short beam pulses, but this kind of time structure is not desirable when the pile-up of very close in time events is a limiting factor, like in fixed-target annihilation experiments (as PADME). In a beam extracted from a RF (warm) LINAC, this parameter is limited to $<10^{-3}$ (e.g. considering <ms long pulses with a repetition rate up to <kHz): for the DAΦNE LINAC it is actually limited to $10^{-5}$ (200 ns/20 ms), it can be increased to $2\cdot10^{-4}$ in case acceleration at half energy (<300 MeV) using the full flat klystron RF power (4 μs/20 ms) is used.

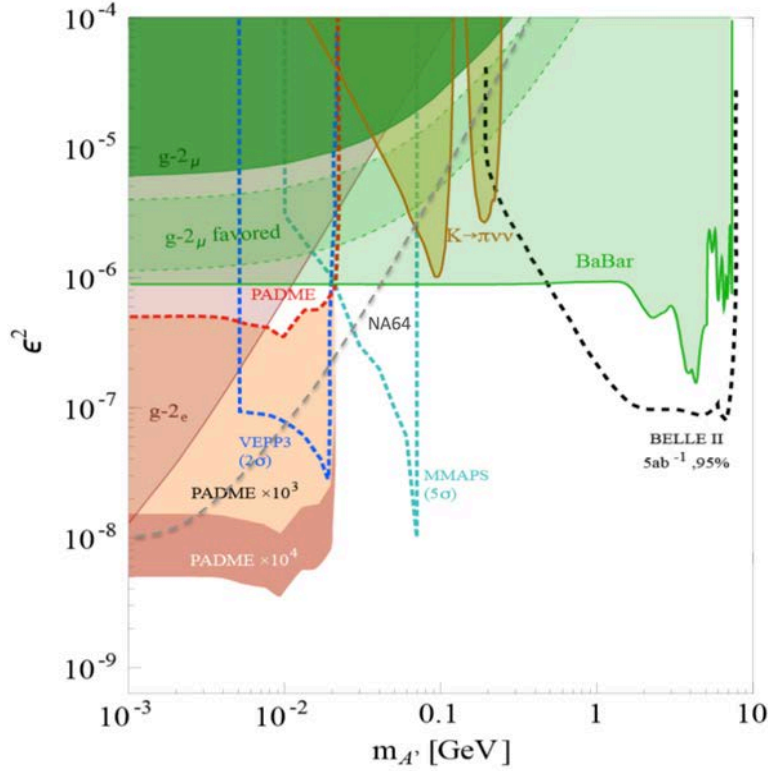

**Figure 7.** *Sensitivity of experimental searches of a dark photon decaying to invisible final states in the plane coupling vs. mass. Present exclusions, and projected sensitivity for future experiments (Belle-II and MMAPS proposal at Cornell) are shown (see text).*

Assuming a $10^{-2}$–$10^{-1}$ range for the duty cycle, i.e. 0.2–2 ms at 50 Hz, the increase of luminosity of a PADME-type fixed-target experiment for light dark matter searches in positron annihilation would be between 3 and 4 orders of magnitude with respect to the running at the DAΦNE BTF. A conservative potential sensitivity of a improved dark photon experiment, using the same PADME approach with such an improved statistics, and without considering the improvement due to the better recoil photon resolution on the background rejection (thanks to the superior beam quality achievable with the slow extraction), is shown in Fig. 7, in comparison with present limits (the main one being the search in single-photon events at BaBar [26]) and other positron annihilation proposed experiments (VEPP-3 at Novosibirsk [21] and MMAPS at Cornell [27], as well as a projection for BELLE-II at the super-KEKB factory [7]).

Indeed further improvement with respect to the PADME experiment as performed with the LINAC positron beam extracted to the BTF line, is expected thanks to the better definition of the positrons spot and divergence, which should allow achieving a better missing mass resolution: this pushes the sensitivity since it reduces the number of background events expected in each bin. As it can be deduced from the plot shown in Fig. 7, the largest room for improvement in sensitivity is at higher dark photon masses, where the dark photon production cross section is enhanced.

The limit at 550 MeV for the positron energy implies a kinematical limit of 24 MeV/$c^2$, but the availability of long positron pulses would also open a more futuristic option for dark sector searches in $e+e-$ annihilations, that is a very asymmetric collider [28]: colliding high-energy positrons with a lower energy ($E_-$) electron beam produced by a small, high duty-factor LINAC, would indeed increase the center of mass energy ($M^2<2E_-E_+$), greatly extending the accessible mass range. However, this option requires most probably a low-energy, cold machine, like an energy-recovery (ERL) super-conducting LINAC, in order to get a significant duty-factor of the collisions together with the very small beam size necessary to get a significant luminosity, and should be more deeply investigated.

## *2.2 Resonant extraction*

Electron beams have been extracted by circulating machines in fewer cases with respect to protons, like the 250 MeV MAX I electron stretcher at MAX-Lab [29], the 5 GeV electrons/positrons extracted from the DORIS ring at DESY [30], the 0.5-3.2 GeV ELSA electron stretcher ring in Bonn [31] and the 250 MeV Saskatchewan electron pulse stretcher (EROS) [32]. Moreover, very few external positron beams are available, like the secondary positrons in the H4 line at the CERN North Area (from the SPS protons), and of course the DAΦNE BTF.

Resonant extraction is a very effective method for extracting particles from a synchrotron, in a relatively large number of turns, based on the simple concept of making them unstable in a controlled way, so that they are driven towards the external part of the beam-pipe, where they are kicked out by means of electrostatic and magnetic fields. The most common choice is a resonance corresponding to a tune value close to a 1/3 of integer, excited by means of a sextupole magnet. A less usual choice for the tune resonance is the half integer, which however produces shorter extraction times, or 1/4 of integer. An extraction method making use of islands of stability, trapping particles and then driving them towards the outside of the horizontal phase space by means of non-linear elements has been recently proposed [33]. Higher orders are not used because of too close separatrices or no unstable area.

In Fig. 8 the deformation of the horizontal phase space ($x,x'$) – which is elliptically shaped area (circular in the normalized coordinates) in a linear machine – is shown for a tune close to 1/2 and 1/3 of integer resonance: the region of stable motion is delimited by curves (green lines), called separatrices. In the case of third order resonance the stability region is thus delimited by a triangle: particles out of the borders of the triangle become unstable and move outward, along the separatrices. The stable triangle does not change area when moving along the machine at a position *s* from the sextupole, it will just get rotated (clockwise) by the betatron phase advance Δμ(*s*).

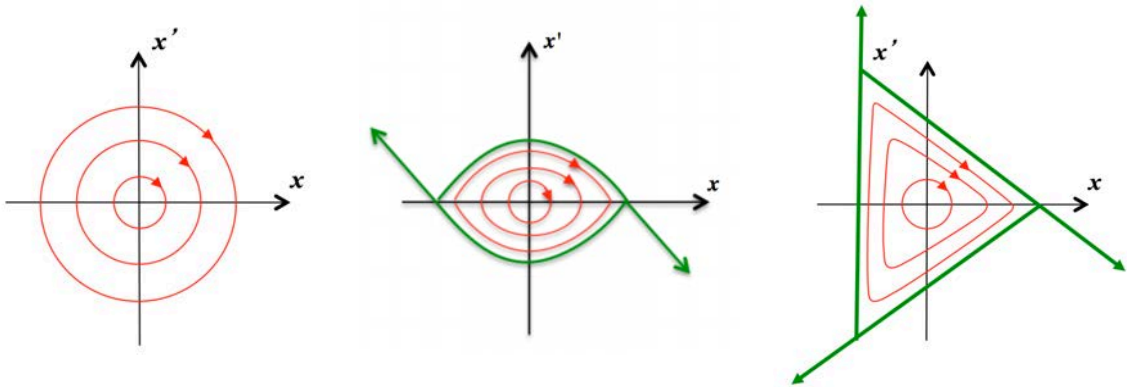

**Figure 8.** *Phase space distributions (x,x') for a linear machine (left) and in case of tune close to 1/2 (middle) and 1/3 of integer resonance (right, δQ>0) at the sextupole. If the tune is below the resonance the rotation in phase space is reversed, if the strength of the sextupole is reversed the triangle is mirrored with respect to the x' axis.*

Exactly at resonance no stable trajectory in the horizontal phase space ($x,x'$) exists, so that by moving the betatron tune towards the resonance, the stability region shrinks and particles with smaller amplitudes become unstable, as illustrated in Fig. 9, reproduced from Ref. [27], where for example the resonant slow extraction at the Cornell University is proposed (5.3 GeV positrons, soon upgraded to 6 GeV, generally injected in the CESR synchrotron light facility, CHESS).

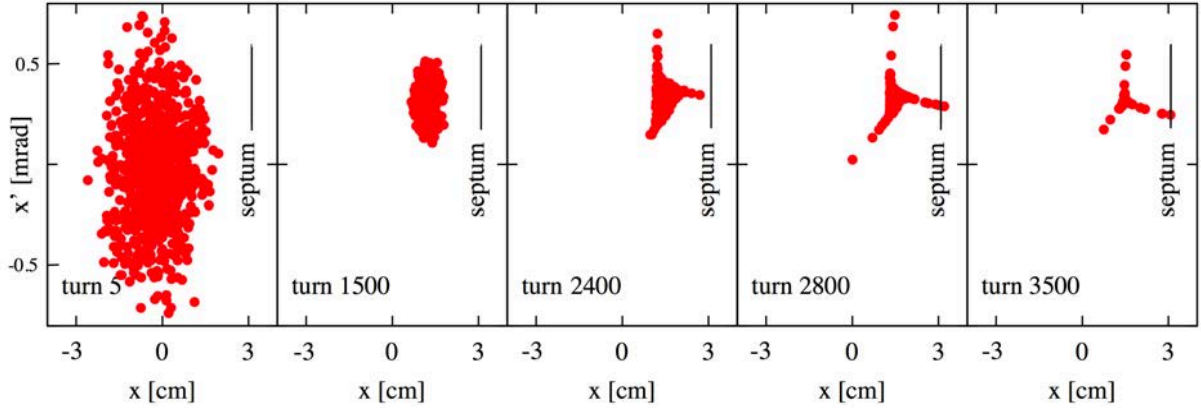

**Figure 9.** *Beam phase space at five points during the acceleration and extraction cycle, illustrating the third order resonance extraction from the Cornell synchrotron (reproduced from Ref. [27]).*

## *2.3 Monochromatic extraction*

A variant of the third integer extraction has been proposed and implemented for the Saskatchewan pulse stretcher ring (EROS) [32], coupling the horizontal betatron tune to the particle energy: if the chromaticity of the ring is non-vanishing, any energy variation will be related to a variation of the tune: $\delta\nu = \chi \cdot \delta p/p$, so that the required variation for moving particles towards the resonance value can be induced by any energy loss mechanism, without making use of quadrupoles and sextupoles. This gives the great advantage of avoiding the use of pulsed magnets. The obvious energy loss mechanism is the synchrotron radiation emission, when the RF of the ring is kept off.

In addition, if the RF is kept off, the LINAC beam pulse can be extended up to filling continuously the entire ring. The beam will be injected in the closed orbit vertically, but off the orbit horizontally: due to the energy spread of the beam and the high chromaticity, the beam will quickly fill an hollow circle in the horizontal phase space, defined by a minimum and maximum emittance value.

An alternative method studied at EROS is to keep the RF on, but making use of a modulation of the frequency or amplitude, or a shift of the phase, for controlling the spilling of particles out of the bucket: adjusting the rate of particles leaving the bucket and being extracted by the synchrotron energy loss mechanism, should allow to optimize the extraction time with respect to the LINAC repetition rate.

## *2.4 The DAΦNE machine*

The DAΦNE electron-positron collider has been designed for reaching a very ambitious target luminosity of $5 \cdot 10^{32}$ cm$^{-2}$s$^{-1}$ at the relatively low center of mass energy of 1 GeV. In order to achieve such high luminosity, the choice has been to have separated electron and positrons storage rings and many bunches (120), very closely spaced along the ~97 m long machine, reaching currents in excess of ~2 A and with very squeezed beams at the interaction point.

The available charge in a single LINAC pulse is of ~1 nC, and in order to improve the emittance of order ~$10^{-5}$ m·rad prior to injecting in the DAΦNE main ring, positrons are stacked and damped in a smaller ring (damping or accumulator ring, exactly 1/3 of the main rings length). In this configuration, the pulse width is ~10 ns at the injection into the accumulator, from where positrons are extracted and injected at 2 Hz in the main ring. Up to 120 bunches circulate in the DAFNE rings, running at an RF of 368.26 MHz, i.e. with a separation of 2.7 ns and a RMS length increasing with the current from <1 cm up to ~3 cm. The same procedure is applied for filling the electron ring.

Since 2008 the large Piwinsky angle or crabbed waist collisions scheme was implemented, allowing to reach a luminosity of >$4 \cdot 10^{32}$ cm$^{-2}$s$^{-1}$, based on increasing the horizontal crossing angle (reducing the horizontal beam size at the interaction point), and suppressing synchro-betatron oscillations by means of sextupoles [34].

The full magnetic layout of the rings with only one interaction region is shown in Fig. 10: electron (blue) and positron (red) rings cross in the middle of two ~10 m long straight sections, where the two interaction points are placed: IP1, used for KLOE and SIDDHARTHA experiments, and IP2 used for the FINUDA experiment. The latter has been replaced with a vertical separation of the two beams and a quadrupole triplet, for implementing the crabbed waist scheme. There are a short and a long arc for each ring, each with four 45° bending dipoles. Arc halves – each containing two dipoles and one wiggler – are connected by short straight sections where the radio-frequency cavities (short arc) and the injection septa (long arc) are placed.

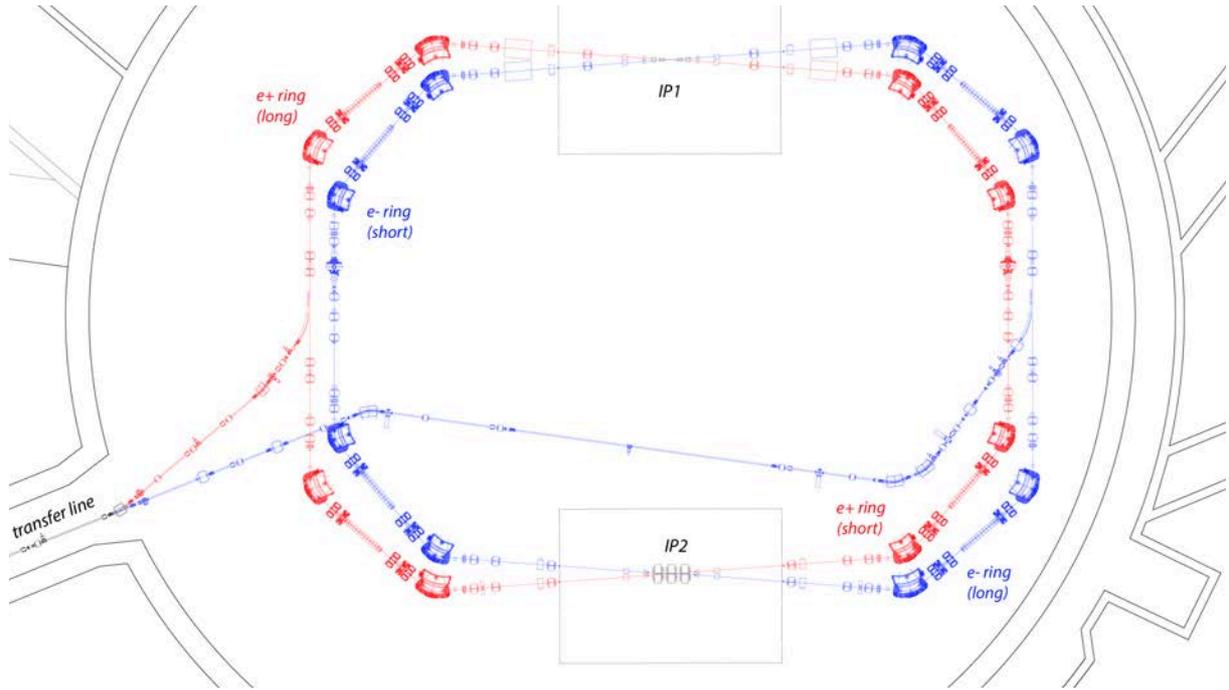

**Figure 10.** *DAΦNE rings magnetic layout with upgraded IP1 interaction region for the crabbed waist operation and beam separation at IP2.*

There are other important peculiarities of the machine, like the large chromaticity due to the interaction region quadrupoles, the damping wigglers (four per ring, where half of the synchrotron radiation power is emitted), and the experiment solenoid and compensators, which have a strong impact on the beams. These elements are strongly connected to the crab waist scheme, adopted for the operation of DAΦNE as high-luminosity collider and would not be used for the resonant extraction scheme.

## 2.5 Slow extraction from DAΦNE

As described above, making use of the monochromatic 1/3 of integer resonant extraction allows stretching positron pulses accelerated by the LINAC, injected directly in one of the DAΦNE rings (the "positron" one), i.e. without being stacked and damped in the accumulator ring. The beam, extracted by suitable septa, is then driven to an external beam-line realized in the DAΦNE hall. Avoiding the injection and extraction in the single bunch intermediate ring has at least a two-fold advantage:

- It allows injecting at a higher frequency than 2 Hz, up to the maximum LINAC repetition rate of 50 Hz (49 effective, since 1 pulse per second is used for energy measurement), thus yielding a higher duty cycle;
- It allows starting from LINAC macro-bunches much longer than 10 ns generally used for DAΦNE operations, being 13.4 ns the maximum longitudinal acceptance of the damping ring (using 1/5 of the main ring RF).

Another interesting point is that in case of upgrade of the LINAC maximum energy the bending dipoles of the DAFNE main rings could be raised from 510 MeV to ~750 MeV, while the smaller accumulator ones are closer to the saturation limit.

There are of course other consequences when choosing this scheme:

- The positron beam emittance is at best the one produced at the end of the acceleration by the LINAC of ~$10^{-5}$ m·rad , i.e. at least an order of magnitude worse with respect to the one at the exit of the accumulator;
- The energy spread of long pulses is increased from 0.5% to a maximum of ~2% (determined by the LINAC and transfer-line acceptance);
- The main ring RF should be kept off, in order to avoid bunching at 368 MHz of the long macro-pulses from the LINAC.

Due to the need of some collimation for improving both the emittance and the energy spread, the number of positrons in the macro-pulse effectively injected in the main ring has to be scaled down from the maximum achievable of $5 \cdot 10^9$/pulse The optics of the transfer line, from the end of the LINAC, will need to be matched to the dispersion and beta functions at the positron injection septum. The emittance of the extracted beam, indeed, will depend on the injected beam size.

The extraction time depends inversely on the synchrotron energy loss, so it will decrease as a function of the beam energy, and will be proportional to the total energy spread. In Ref. [35] the main parameters of the extracted beam have been calculated for the DAΦNE, considering as input $5 \cdot 10^7$ positrons in 150 ns long pulses from the LINAC, a horizontal chromaticity of $C_x$=-3, a jump $\Delta x$=5 mm and septum thickness δ =0.1 mm, respectively with the wigglers turned on (off):

- $N_{e+}$ ~$2 \cdot 10^9$ s$^{-1}$;
- Extraction time ~0.5 (0.2) ms;
- Energy spread ~$10^{-3}$;
- Efficiency ~98%;
- Emittance $W_x$~$3 \cdot 10^{-6}$ m·rad, $W_y$ same as injection.

The achievable duty cycle, >0.2 ms/20 ms, is at least a three orders of magnitude improvement with respect to the extraction from the LINAC. Moreover, much better energy spread and emittance are expected. A conceptual layout of the direct injection and the new extraction line inside the DAΦNE hall is shown in Fig. 11.

An electrostatic and a magnetic septum would then be needed to drive the positrons to a dedicated transfer-line for further beam manipulation, and both the interaction regions have to be re-adjusted in order to match the optics, in particular at the IP2; moreover the sextupoles strength has to be computed, the extraction line optics optimized, etc.

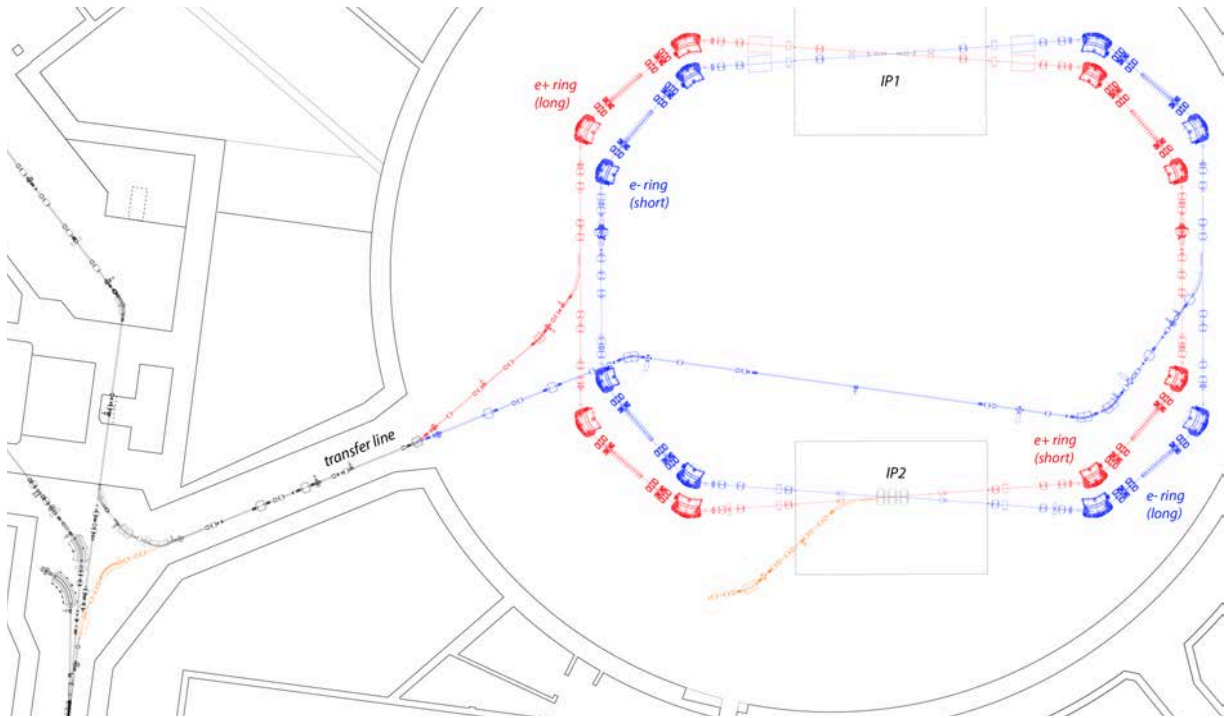

**Figure 11.** *Conceptual layout of the DAFNE main rings (blue=electrons, red=positrons) with modified injection line from LINAC and the positron extraction line (orange) at the IP2 straight section.*

## 2.6 RF control of extraction

The proposed monochromatic extraction scheme does not allow to control the extraction time, and thus the achieved duty cycle, since the synchrotron radiation loss simply scales with the fourth power of the beam energy while the size of the horizontal phase space is proportional to the energy spread. If the energy loss is too small, the extraction will be too slow, and only part of the injected beam will be driven to the septum in the time between two LINAC pulses. This is not the case for DAΦNE typical at 500 MeV, where the 2% energy spread of the injected beam is lost by synchrotron radiation in ~0.5 ms, a time much shorter than the LINAC pulses spacing, yielding a ~$10^{-2}$ duty cycle, as described in the previous section.

In order to control the extraction time, it is possible to restore the energy loss by means of the RF cavity, and to drive the particles towards the resonance by varying some quadrupoles and sextupoles in the ring. An alternative method for shrinking the stable region in the horizontal phase space, while controlling the speed with which circulating particles are spilled out from the stability triangle, is to increase the amplitude of betatron oscillations. This can be achieved by perturbing the beam with a transverse RF electric field, resonating with the horizontal betatron oscillation: this is the so-called RF knockout. This technique is usually used for ion and proton beams, but has been also implemented for the 100 MeV electron ring at KSR (Korea) [36].

In principle, duty-cycle values close to 100% can be reached, as demonstrated in the electron pulse stretcher EROS, but detailed studies are needed for investigating this option for DAΦNE. In any case, the price to be paid is the fact that the RF will bunch the injected beam, resulting in a reduction of the effective duty-cycle, even though this would allow achieving higher values of the extraction time. Moreover with this scheme there is no control on the energy spread of the extracted beam. Also in this case further studies are needed, and possibly a combination of the two techniques could be adopted for DAΦNE, allowing at the same time to increase the duty factor and reduce the energy spread of the extracted beam.

## 2.7 Crystal extraction

A crystal can channel a charged particle if it comes within so-called critical angle $\theta_c$, ~5 μrad/$p$(TeV) for Silicon. Particles in a circulating beam can be thus deflected (by a small angle) when passing (many times) through crystal short enough for reducing scattering losses. Since the deflection angle is small, in this non-resonant technique the crystal is the primary element that has to be followed by a septum magnet like in the resonant schemes.

Non-resonant extraction through deflection of charged particles in bent crystals has been used routinely for proton beams in Protvino (U-70), and also demonstrated at CERN (SPS) and Fermilab (Tevatron), always for multi-GeV energy beams.

At sub-GeV energy, however, the de-channeling length (characteristic dimension over which the crystal efficiently deflects the particles) gets relatively small, thus requiring very thin crystals. This is one reason for which, even though channeling effects have been observed also with electron and positron beams, crystal-assisted extraction has never been used for the extraction from electron rings. On the other side, at very low energies the thickness of the crystal is a crucial parameter in order to achieve multi-turn extraction, since the stored beam has not to be destroyed by the interactions with the material. In any case, also considering the improved crystal fabrication technology and possible alternative solutions to classical bent crystals (like channelling in planar arrays, half-wavelength crystals, etc.) make this option interesting, even though surely challenging.

Finally, in addition to the general scientific interest of such studies, this technique would offer a number of potential advantages, starting from the high collimation of the channelled particles, providing excellent beam quality of the extracted beam.

## 2.8 Second DAΦNE ring

This proposal is mainly centered on the use of the positron ring as a pulse stretcher, but of course the availability of the second ring, together with a relevant research infrastructure which the DAΦNE-light laboratory, with its synchrotron radiation lines, installed on the external arc of the electron ring, naturally suggests the option of keeping in operation also the second ring with a stored beam. The proposed pulse stretcher does not conflict with this possibility, even though different choices will imply some slight modification to the project, e.g. the need of separating the two beams in both interaction points.

Another relevant point is the accumulator or damping ring: the proposed positron monochromatic slow extraction does not rely on the damping ring, so that it would be fully compatible with a re-use of the area and transfer-lines, for instance as a high-radiation electron dump facility, which can have multiple applications:

- Neutron electro-production;
- Electron irradiation;
- High-intensity dump experiments (e.g. dark sector particles searches).

Even though this possibility is not discussed in this document, it would have a significant impact on the required resources, since many (if not all) the needed components can be recuperated by the dismantling of the damping ring.

In case the second ring is used for storing an electron beam, the typical parameters of the LINAC electron beam (~$10^{-6}$ m·rad emittance, 0.5% energy spread) should allow the injection even without the intermediate damping ring. The DAΦNE electron beam could in principle still be used as electron storage ring, just switching to the electron mode (also adjusting the gun parameters: no positron converter, short pulses), and directly injecting in the electron ring a suitable number of pulses for getting the required filling pattern. This will allow storing single or any pattern of electron bunches, both for the synchrotron radiation lines and for machine studies with the electron circulating beam.

The project is also compatible with other options, like the one of storing a high-current positron beam for machine studies, like those on the electron cloud instability. In order to have an acceptable positron injection in the second ring, short pulses should be injected, and in this case keeping in operation the damping ring could be beneficial.

In this hypothesis, the double use of the machine as positron pulse stretcher (in one ring) and positron storage ring (in the other), should not be an issue since the lifetime of the non-colliding stored beam could be long enough to require injections spaced by long periods, thus with a small impact on the extracted beam duty-cycle. Of course such an option should be further studied, for instance in connection with the use of the synchrotron lines, installed on the "electron" ring. In this case running the synchrotron lines with positrons would be still possible, simply reversing the polarity of all magnets of the main ring (and of the electron transfer-line inside the hall).

In any case, if a second beam is injected and stored, the two rings have to be separated, for instance in the vertical plane (as done already in the KLOE-2 configuration, at the IP2).

Finally, in the case in which only positrons would be required for the injection in both rings, this would impact on the LINAC operation, since it would be kept in positron mode, with practically no dead time for reversing the transfer-lines (and setting in and out the positron converter), apart from the switch of the dipole splitting the injection lines, at the end of the common transfer line. On the other hand, the standard long pulse configuration of the LINAC gun needed for the stretcher ring (the "positron" one) should be changed periodically for allowing injection of short pulses in the damping ring for the refilling of the storage ring (the "electron" one).

## 3. Outlook and perspectives

*3.1 Resources*

An accurate estimate of the needed resources can be performed only when some more details of the project will be defined, however already at this stage of the proposal some preliminary considerations are possible in order to make a first, rough estimate of the additional hardware, as well as the entity of the modifications on the accelerator complex. In particular, considering a time horizon >2020 for this kind of project, two main options can be foreseen: dismantling the accumulator or keeping the possibility of using the intermediate ring.

In the first case many of the components, in particular magnetic elements and vacuum pipes can be recuperated. The main new elements will be the extraction septa and possibly additional sextupoles with respect to the ones already installed in DAΦNE. However, as discussed above, dismantling the damping ring would prevent efficient injection of positrons in one of the main rings, which instead would an interesting option even when DAFNE will be no longer running as a collider.

A preliminary list of components is here summarized (some of them could be re-used, depending on the re-arrangement of the main rings and of the transfer-lines):

- *Interaction regions*
    - 4 quadrupoles for IP1: electro-magnets replacing permanent quadrupoles
    - Extraction septa for IP2
    - 2 dipoles ~30°
    - 4-6 quadrupoles and correctors for extraction line
    - Vacuum pipes ~10 m.
- Sextupoles in the zero dispersion machine regions for exciting the resonance, in addition to sextupoles for controlling the chromaticity in dispersive regions.
- *Direct injection line*
    - 15° pulsed dipole
    - 45° DC dipole

- 4 quadrupoles and correctors
- Vacuum pipes
- Collimators.

Some additional resource will be needed in case the crystal extraction will be implemented.

## *3.2 Outlook and future work*

Various studies have to be started in order to define this proposal in some more detail:

- Full design of the third integer resonant extraction, including the modifications to the DAΦNE optics for the optimization of the transport of the excited particles to the septa;
- Design of the electrostatic and magnetic septa, for the optimization of the extracted beam parameters;
- Design of the injection and extraction magnetic lines, with the aim of preserving the extracted positron beam quality;
- Optimization of the positron production at the LINAC, of the collimation and focusing in order to decrease both the emittance and the energy spread, and of the injection into the DAΦNE ring. A good compromise between beam intensity and quality is needed for an optimal direct injection, i.e. without using the intermediate damping ring;
- Study of the monochromatic extraction, tuning of the extraction time through the ring RF and optimization of the duty-cycle;
- Study of the option of non-resonant extraction, with a crystal channelling setup.

The design phase has then to be followed by the modification of the DAΦNE layout in order to arrange the required elements for the resonant extraction, including the septa and the building of the new transfer-line for the handling and delivery of the extracted positron beam.

A preliminary estimate of the time schedule can be done, considering that the complex will be used for collisions at least until the end of 2019 and that the LINAC and BTF facility will also undergo an important consolidation and upgrade program: after one year of studies for a better definition of the project and for preparing a fully detailed technical report, new components and can be designed and procured in the following year, so that the new hardware installation, adaptation of the DAΦNE ring, dismantling of the DAΦNE damping ring and modifications to the injection transfer-line can start in the second half of 2020. The commissioning of the beam and the experimental activities can thus be foreseen in 2021.

## 4. Acknowledgements

The original very schematic idea developed thanks to the discussions with Jim Alexander, Andrea Ghigo, Mauro Raggi, David Rubin and Bogdan Wojtsekhowski.

I am very grateful for the suggestions and indications to Susanna Guiducci, who also performed the first calculations for the extraction from DAΦNE [35].

Finally, thanks to Tommaso Spadaro and Mikhail Zobov for the critical review of the text.

This work is supported by the Italian Ministry of Foreign Affairs and International Cooperation (MAECI), CUP I86D16000060005.